# A Statistical Fat-Tail Test of Predicting Regulatory Regions in the *Drosophila* Genome

**Jian-Jun SHU*** and **Yajing LI**
School of Mechanical & Aerospace Engineering, Nanyang Technological University,
50 Nanyang Avenue, Singapore 639798.

## ABSTRACT

A statistical study of *cis*-regulatory modules (CRMs) is presented based on the estimation of similar-word set distribution. It is observed that CRMs tend to have a fat-tail distribution. A new statistical fat-tail test with two kurtosis-based fatness coefficients is proposed to distinguish CRMs from non-CRMs. As compared with the existing fluffy-tail test, the first fatness coefficient is designed to reduce computational time, making the novel fat-tail test very suitable for long sequences and large database analysis in the post-genome time and the second one to improve separation accuracy between CRMs and non-CRMs. These two fatness coefficients may be served as valuable filtering indexes to predict CRMs experimentally.

## 1. INTRODUCTION

The identification of transcription factor binding sites (TFBSs) and *cis*-regulatory modules (CRMs) is a crucial step in studying gene regulation. Computational methods of predicting CRMs can be classified into three types: 1) TFBS-based methods, 2) homology-based methods and 3) content-based methods. TFBS-based methods, such as ClusterBuster (Frith, *et al.*, 2003) and MCAST (Bailey and Noble, 2003), use information about known TFBSs to identify potential CRMs. This type of approach cannot be applied to genes for which TFBSs have not been experimentally studied. Homology-based methods use information contained in the pattern of conservation among related sequences. The related sequences can come from single species (van Helden, *et al.*, 1998), two species (Grad, *et al.*, 2004) and multiple species (Boffelli, *et al.*, 2003). The performance of this approach using only conserved patterns is limited because the conservation of TFBS required to maintain regulatory function may not be significantly higher in binding sequences than in non-binding sequences (Emberly, *et al.*, 2003). In addition, it still remains an open question how many genomes are sufficient to the reliable extraction of regulatory regions. Content-based methods assume that different genomic regions (CRMs, exons, and NCNRs) have different rates of evolutionary microscopic change; therefore, they exhibit different statistical properties in terms of nucleotide composition. TFBSs often occur together in clusters as CRMs (Berman, *et al.*, 2002; Lifanov, *et al.*, 2003). The binding site cluster causes a biased word distribution within CRMs, and this bias leaves a distinct "signature" in nucleotide composition. Content-based methods detect this signature through statistical techniques (Nazina and Papatsenko, 2003; Abnizova, *et al.*, 2005) or machine learning techniques (Chan and Kibler, 2005) to distinguish CRMs from non-CRMs. Such methods can be used to predict CRMs that have not been observed experimentally. A large number of CRM search tools have been reported in the literature, but computational method attempting to identify CRMs still remains a challenging problem due to the limited knowledge of specific interactions involved (Su, *et al.*, 2010; Shu and Li, 2013).

* Correspondence should be addressed to Jian-Jun SHU, mjjshu@ntu.edu.sg

The fluffy-tail test (Abnizova, *et al.*, 2005) is one of content-based methods. It is a bootstrapping procedure to identify CRMs by checking the statistical difference between the size distribution of the largest group of similar-words obtained for the randomized shuffled sequences and the corresponding size distribution for the original input nucleotide sequence. If there are no statistical differences, it is concluded that the original input nucleotide sequence is a coding (exon) region or a non-coding non-regulatory (NCNR) region.

In the work that follows, the fluffy-tail test is re-examined by considering the following two issues: 1) Due to its bootstrapping procedure, the computational time of calculating the fluffiness coefficient is determined by the number of randomizations. To obtain statistically reliable results, the number of randomizations in the fluffy-tail test is usually set to be exceptionally large, and thus expensive to compute, especially for long sequences. This limits the use of the fluffy-tail test under the situation when increased DNA sequences need to be analyzed in the post-genome time. 2) The fluffy-tail test looks only at the subsequence with the highest incidence in the CRMs. Therefore, the fluffy-tail test may not capture the statistical features caused by heterotypic TFBS clusters in the regulatory regions (Shu and Li, 2013). It is an interest to address these two issues of the fluffy-tail test and to develop a more efficient and effective CRM prediction method.

This paper is to explore some statistical properties of DNA composition due to the multiple occurrences of TFBSs of the same or diverse types in CRMs. For an enumeration purpose, a consensus sequence is used as a motif representation, *i.e.*, using a similar-word set to represent a motif. The main concern is to explore specific properties in the similar-word set distribution for CRMs and identify suitable parameters to distinguish CRMs from non-CRMs.

## 2. MATERIALS AND METHODS

### 2.1 Training datasets

To explore statistical parameters to distinguish CRMs from non-CRMs, three training datasets are used in this paper. The positive training set is a collection of 60 experimentally-verified functional *Drosophila melanogaster* regulatory regions (Papatsenko, *et al.*, 2002; Nazina and Papatsenko, 2003). This set consists of CRMs located far from gene coding regions and transcription start sites. It contains many binding sites and site clusters, including *abdominal-b*, *bicoid*, *caudal*, *deformed*, *distal-less*, *engrailed*, *even-skipped*, *fushi tarazu*, *giant*, *hairy*, *huckebein*, *hunchback*, *knirps*, *krüppel*, *odd-paired*, *pleiohomeotic*, *runt*, *tailless*, *tramtrack*, *twist*, *wingless* and *zeste*. The total size of positive training sets comprises about 99 kilobase (kb) sequences. The two negative training sets are: 1) 60 randomly-picked *Drosophila melanogaster* exons, and 2) 60 randomly-picked *Drosophila melanogaster* NCNRs: The exons and NCNRs of length 1 kb upstream and downstream of genes are excluded by using the Ensembl genome browser. The exon training set contains 85 kb sequences, and the NCNR training set contains 90 kb sequences. All sequences with tandem repeats in the three training datasets are masked by using a tandem repeats finder program (Benson, 1999) before processing.

## 2.2 Formulation of the fat-tail test

The fat-tail test assumes that each word (binding site) recognized by a given transcription factor belongs to its own family of similar-word sets (binding site motifs) found in the same enhancer sequence and the redundancy of binding sites within CRMs leaves distinct "signatures" in similar-word set distribution. For a given $m$-letter segment $W_m$ as a seed-word, all $m$-letter words that differ from $W_m$ by no more than $j$ substitution comprise a corresponding similar-word set $N_j(W_m)$. Because the core of TFBSs is relatively short (Zhu, *et al.*, 2011), a five-letter seed-word is selected, allowing for one mismatch, that is, $m = 5$ and $j = 1$. The fat-tail test is adopted to study the similar-word set distribution and to predict the probable function of the original input sequence. A flow chart of the fat-tail test is shown in Figure 1.

<u>Step 1: Number of similar-words with the same seed-word</u> ($n$)
As an example, consider a stretch of DNA: ACGACGCCGACT. For $m = 5$ and $j = 1$, all five-letter segment $W_5$ is selected as a seed-word, that is, ACGAC , CGACG , … , CGACT , … . For each seed-word $W_m$, all $m$-letter words with no more than $j$ substitution comprise a corresponding similar-word set $N_j(W_m)$. In this example, the first seed-word $W_5$, ACGAC, has three similar-words with no more than one mismatch: ACGAC , ACGCC , CCGAC, $n$ is the cardinality, $n = |N_j(W_m)| = |N_1(\text{ACGAC})| = 3$, and forms X axis in Figures 2-7.

<u>Step 2: Number of seed-words with the same number of similar-words</u> ($f$)
$f(n)$ is the number of seed-words containing $n$ similar-words and forms Y axis in Figures 2-9.

<u>Step 3: Kurtosis</u> ($k$)
The kurtosis $k$ of similar-word set distribution $f(n)$ is evaluated as

$$k = \frac{\sum_{n=1}^{N} [f(n) - \mu]^4}{(N-1)\sigma^4} - 3 \tag{1}$$

where $\mu$ and $\sigma$ are the mean and standard deviation, respectively.

<u>Step 4: Two fatness coefficients</u> ($D$ <u>and</u> $S_r$)

The first fatness coefficient $D$ is defined as:

$$D = \frac{k_0 + 2\varepsilon}{4\varepsilon}. \tag{2}$$

Here $k_0$ denotes the kurtosis $k$ of the original input sequence without randomly-shuffling and $\varepsilon$ is the standard error calculated by:

$$\varepsilon = 2\sqrt{\frac{6}{N}}. \tag{3}$$

$D$ is used to measure how strongly the similar-word set distribution of CRMs deviates from normal distribution. The 95% confidence interval is set between $-2\varepsilon$ and $2\varepsilon$. To measure how strongly the similar-word set distribution of CRMs deviate from randomness, the second fatness coefficient $S_r$ is computed by comparing with all randomized $r$-time shuffled sequence versions of the original input sequence:

$$S_r = \frac{k_0 - k_r}{\sigma_r} \quad (4)$$

Here a sequence is called "random" if it is obtained from the original input sequence by shuffling it, preserving its single nucleotide composition. $S_r$ can be regarded as measuring the degree of difference between signal and noise, where the signal is regarded as the original input sequence, and the noise is regarded as randomized sequences. In the fluffy-tail test (Abnizova, *et al*., 2005), the fluffiness coefficient $F_r$ is defined as:

$$F_r = \frac{L_0 - L_r}{\sigma_r} \quad (5)$$

where $L_r$ is the number of seed-words with the maximal similar-words for $r$-time shuffled sequences. Here it is worth to mention to this end that CRMs tend to have a fat-tail distribution in Figure 2, as compared with the randomized sequence in Figure 3. Since kurtosis measures the tail heaviness of a distribution relative to a normal distribution, the second fatness coefficient $S_r$ based on kurtosis $k_r$ should be a more reasonable predictor of CRMs than the fluffiness coefficient $F_r$ based on the maximal number $L_r$.

# 3. RESULTS

## 3.1 Distribution for CRMs

For the training datasets of CRMs, Figure 2 shows a similar-word set distribution for a region of *Drosophila melanogaster hunchback* CRMs. It can be seen that the most frequent similar-word set occurs 10 to 40 times and some similar-word sets occur about 95 times. If the original input sequence is characterized by the presence of an unusually-high number of over-represented similar-words, then the similar-word set distribution is expected to have a longer right tail than a random sequence, since $(k_0 = 4.19)$ is much larger than $(k = 0)$ for the normal distribution.

To obtain a random distribution, the original input sequence is shuffled 50 times by using the Fisher–Yates shuffle algorithm. Figure 3 shows a typical example of similar-word set distribution after randomly-shuffling. Compared with the original input sequence in Figure 2, the randomized sequence in Figure 3 lacks a long right tail and is close to a normal distribution, in view of $(k_r = 0.19)$ around 0.

## 3.2 Distribution for exons

For the training datasets of randomly-picked *Drosophila melanogaster* exons, Figure 4 shows a similar-word set distribution for a region of *Drosophila melanogaster CG8229* exons. As can be seen in Figure 4, there is no long right tail, in view of that $(k_0 = -0.28)$ is around 0. Figure 5 shows a typical example of similar-word set distribution after randomly-shuffling with $(k_r = 0.35)$ around 0. The kurtosis $k_0$ of similar-word set distribution for the original input sequence doesn't differ significantly from $k_r$ of the randomized version, $(k_0 = -0.28)$ *vs.* $(k_r = 0.35)$.

### 3.3 Distribution for NCNRs

For the training datasets of randomly-picked *Drosophila melanogaster* NCNRs, Figure 6 shows a similar-word set distribution for a region of *Drosophila melanogaster* NCNRs. The presence of the short right tail is noticed in Figure 6 in view of that $(k_0 = 0.09)$ is around 0. Figure 7 shows a typical example of similar-word set distribution after randomly-shuffling with $(k_r = 0.25)$ around 0. The kurtosis $k_0$ of similar-word set distribution for the original input sequence doesn't differ significantly from $k_r$ of the randomized version, $(k_0 = 0.09)$ *vs.* $(k_r = 0.25)$.

### 3.4 The fat-tail test

To distinguish CRMs from non-CRMs, $D$ and $S_r$ are calculated on 180 sequences in three training datasets. Figure 8 shows that CRMs tend to have a greater $D$ than exons and NCNRs. Table 1(a) lists functional classification based on $D$. 75% CRMs have $D > 2$, while only 18.3% exons have $D > 2$, and 53.3% NCNRs have $D > 2$. Figure 9 shows $S_{50}$ for CRMs, exons and NCNRs. For each sequence, its $(r = 50)$-time shuffled versions are generated to calculate $S_{50}$. It can be seen that CRMs intend to have a greater $S_{50}$ than exons and NCNRs. Table 1(b) lists functional classification based on $S_{50}$. 76.7% CRMs have $S_{50} > 2$, while only 11.7% exons have $S_{50} > 2$, and 36.7% NCNRs have $S_{50} > 2$.

### 3.5 Large CRM datasets

The fat-tail algorithm has been evaluated on the current version 3 of *REDfly* database (Gallo, *et al*., 2011), which contains 894 experimentally-verified CRMs from *Drosophila*. Results show that 63.1% CRMs have $D > 2$ and 59.5% CRMs have $S_{50} > 2$ passing the fat-tail test. The low pass rate may be due to the stringent threshold value. Another probable reason is that some CRMs do not contain binding site cluster. This directs future study: (1) to check if the binding site clustering is the common feature of all CRMs; (2) to optimize the threshold to get more reliable results. It is worth to mention to the point that the fluffy-tail algorithm has never been evaluated on the large CRM datasets.

## 4. DISCUSSION

Some statistical properties of similar-word set distribution in three training datasets have been explored. The results show that CRMs have a fat-tail distribution, *i.e.*, tend to have higher fatness coefficients $(D > 2, S_r > 2)$, whereas exons lack a fat-tail distribution, *i.e.*, tend to have lower fatness coefficients; however, NCNRs tend to have median fatness coefficients. Thus, $D$ and $S_r$ can be used to distinguish between CRMs and exons effectively. CRMs are predominant if $(D > 2, S_r > 2)$, while exons are prevailing if $(D < 2, S_r < 2)$. Thus, the regions with $(D > 2, S_r > 2)$ are CRMs and those with $(D < 2, S_r < 2)$ are exons.

### 4.1 Comparison with the fluffy-tail test

The fat-tail test is evaluated by comparison with the fluffy-tail test (Abnizova, *et al.*, 2005). The performance of three parameters is assessed: 1) the first fatness coefficient $D$, 2) the second fatness coefficient $S_r$ and 3) the fluffiness coefficient $F_r$ based on separation between CRMs and exons, and between CRMs and NCNRs.

The training datasets are employed to evaluate the above three parameters. For comparison, the original input sequence is shuffled 50 times to calculate $S_{50}$ and $F_{50}$. The thresholds of $D$, $S_{50}$ and $F_{50}$ are all set as 2. For the fat-tail test, the original input DNA sequence is considered with $D > 2$ as predicted CRMs, $D < 2$ as predicted exons, and $S_{50} > 2$ as predicted CRMs, $S_{50} < 2$ as predicted exons. For the fluffy-tail test, the original input DNA sequence is considered with $F_{50} > 2$ as predicted CRMs, $F_{50} < 2$ as predicted exons. The classification result of 180 sequences in the training datasets by $F_{50}$ is listed in Table 1(c). The fluffy-tail test $F_{50}$ identified 42 out of 60 CRMs in the positive training datasets, while the fat-tail test identified 45 and 46 CRMs with $D$ and $S_{50}$, respectively (see Table 1). For each parameter, sensitivity (SN) (number of true positive/number of positive), specificity (SP) (number of true negative/number of negative) and accuracy (number of true positive+number of true negative)/(number of positive+number of negative) are calculated to distinguish CRMs from exons and NCNRs (Table 2).

For distinguishing CRMs from exons, the fat-tail test with $S_{50}$ has the best accuracy (82.5%), as compared with the other two parameters ($D$: 78.3%; $F_{50}$: 78.3%). Thus, the fat-tail test with $S_{50}$ can effectively distinguish between CRMs and exons. Moreover, $S_{50}$ (SN=76.7%) can more efficiently identify CRMs than $D$ (SN=75%) and $F_{50}$ (SN=70%), as well as $S_{50}$ (SP=88.3%) can more efficiently identify exons than $F_{50}$ (SP=86.7%) and $D$ (SP=81.7%). The fat-tail test with $D$ has the same accuracy as the fluffy-tail test; however, the computational time (CPU time) to calculate $D$ for the original input DNA sequence of length 1000 is 50 times faster than those to calculate $F_{50}$ and $S_{50}$ for the same original input sequence, because of no 50-time randomly-shuffling required for calculating $D$. Thus, the fat-tail test with $D$ is very suitable for long sequences and large database. For distinguishing CRMs from NCNRs, the results show that the accuracy (67.5%) of the fluffy-tail test with $F_{50}$ is worse than (70%) of the fat-tail test with $S_{50}$, but better than (60.8%) of the fat-tail test with $D$.

### 4.2 Time complexity

Table 3 shows that the value of the fat-tail kurtosis coefficient $S_r$ is affected by the number of randomizations, $r$. To obtain a more reliable estimation of $S_r$, a large $r$ is needed, so that longer computational time is expected. In order to obtain reliable results in a reasonable computational time, the original input sequence is shuffled by 50 times to calculate $S_r$.

The algorithm used for shuffling is the Fisher–Yates shuffle algorithm, which is linear on the sequence length $N$, so that the time complexity of calculating $D$ is $O(N)$ and the time complexity of calculating $S_r$ and $F_r$ is $O(Nr)$. In Table 2(c), the computational time (CPU time) of calculating $D$ is 50 times faster than those of calculating $F_{50}$ and $S_{50}$ due to no sequence shuffling. All computations are run on a 3.2 GHz Pentium IV processor with 1G physical memory.

### 4.3 Tandem repeat region

The results show that the most frequent similar-word set usually corresponds to the word of 'TTTTT' or 'AAAAA' for CRMs and NCNRs. These phenomena are due to the poly N (such as TTT…) occurrence in CRMs and NCNRs and affect the maximal number $L_r$. Thus, true CRMs cannot be distinguished from NCNRs effectively in the fluffy-tail test. The motifs corresponding to experimentally-verified TFBSs usually occur more than the mean value of similar-word set distribution and locate around the right tail, so that the prediction accuracy using the kurtosis-based fatness coefficient $S_r$ is improved. It is worth to mention to this end that the phenomenon of motif fat-tail distribution can be also observed in protein sequences (Bastien, *et al*., 2004; Bastien, 2008; Bastien and Marechal, 2008; Comet, *et al*., 1999; Shu and Ouw, 2004; Shu and Li, 2010; Shu, *et al*., 2012; Shu and Yong, 2013; 2017).

## 5. CONCLUSION

The redundancy of binding sites within CRMs causes the bias base composition and leaves distinct "signatures" in similar-word set distribution. The fluffy-tail test captured this characteristic by searching for the most common similar-words; however, the true binding site motif is likely to be the moderate set of similar-words. In this paper, the fat-tail test is proposed to distinguish CRMs from non-CRMs. In the fat-tail test, using a dataset of 180 DNA sequences (60 for CRMs, 60 for exons and 60 for NCNRs), characteristics are investigated by examining distribution patterns. Results show that the similar-word set distribution of CRMs tends to be a fat-tail distribution as compared with those of exons and NCNRs. Based on this observation, two kurtosis-based fatness coefficients $D$ and $S_r$ are introduced here. The fat-tail test with $D$ has comparable accuracy to, but $r$ times faster than the fluffy-tail test, because of no $r$-time randomly-shuffling required. The fat-tail test with $S_r$ has better accuracy in distinguishing CRMs from exons and NCNRs than the fluffy-tail test. Thus, the novel fat-tail test simplifies the functional prediction of an original input DNA sequence and can guide future experiments aimed at finding new CRMs in the post-genome time (Shu, *et al*., 2011; 2015; Wong, *et al*., 2015; Shu, 2017; Shu, *et al*., 2017).

## Table Captions

Table 1: Classification of 180 sequences

Table 2: Evaluation of $D$, $S_{50}$ and $F_{50}$

Table 3: Sensitivity of $S_r$ to choice of $r$ for CRMs $(k = 4.19)$

**Table 1.** Classification of 180 sequences

(a) Fat-tail test with $D$

| Functional type | $D > 2$ | $D < 2$ | Positive rate | Negative rate |
|---|---|---|---|---|
| CRMs | 45 | 15 | 75% | 25% |
| Exons | 11 | 49 | 18.3% | 81.7% |
| NCNRs | 32 | 28 | 53.3% | 46.7% |

(b) Fat-tail test with $S_{50}$

| Functional type | $S_{50} > 2$ | $S_{50} < 2$ | Positive rate | Negative rate |
|---|---|---|---|---|
| CRMs | 46 | 14 | 76.7% | 23.3% |
| Exons | 7 | 53 | 11.7% | 88.3% |
| NCNRs | 22 | 38 | 36.7% | 63.3% |

(c) Fluffy-tail test

| Functional type | $F_{50} > 2$ | $F_{50} < 2$ | Positive rate | Negative rate |
|---|---|---|---|---|
| CRMs | 42 | 18 | 70% | 30% |
| Exons | 8 | 52 | 13.3% | 86.7% |
| NCNRs | 21 | 39 | 35% | 65% |

**Table 2.** Evaluation of $D$, $S_{50}$ and $F_{50}$

(a) Distinguishing CRMs from exons

| | Fat-tail test | | Fluffy-tail test |
|---|---|---|---|
| | $D$ | $S_{50}$ | $F_{50}$ |
| SN | 75% | 76.7% | 70% |
| SP | 81.7% | 88.3% | 86.7% |
| Accuracy | 78.3% | 82.5% | 78.3% |

(b) Distinguishing CRMs from NCNRs

| | Fat-tail test | | Fluffy-tail test |
|---|---|---|---|
| | $D$ | $S_{50}$ | $F_{50}$ |
| SN | 75% | 76.6% | 70% |
| SP | 46.7% | 63.3% | 65% |
| Accuracy | 60.8% | 70% | 67.5% |

(c) CPU time for a sequence length of 1000

| | Fat-tail test | | Fluffy-tail test |
|---|---|---|---|
| | $D$ | $S_{50}$ | $F_{50}$ |
| CPU time | 6.2 second | 310 second | 310 second |

**Table 3.** Sensitivity of $S_r$ to choice of $r$ for CRMs $(k = 4.19)$

| $r$ | $S_r$ | $k_r$ | $\sigma_r$ |
|---|---|---|---|
| 50 | 3.63 | 0.26 | 0.67 |
| 100 | 5.31 | 0.2 | 0.47 |
| 500 | 4.28 | 0.2 | 0.58 |

**Figure Captions**

Figure 1: A flow chart of fat-tail test

Figure 2: Histogram of *Drosophila* CRMs $(m = 5, j = 1, k = 4.19, \mu = 24.4, \sigma = 11.7)$

Figure 3: Histogram of *Drosophila* CRMs $(m = 5, j = 1, k = 0.19, \mu = 23.9, \sigma = 7.7)$ after randomly-shuffling

Figure 4: Histogram of *Drosophila* exons $(m = 5, j = 1, k = -0.28, \mu = 21.73, \sigma = 7.33)$

Figure 5: Histogram of *Drosophila* exons $(m = 5, j = 1, k = 0.35, \mu = 21.4, \sigma = 7.19)$ after randomly-shuffling

Figure 6: Histogram of *Drosophila* NCNRs $(m = 5, j = 1, k = 0.09, \mu = 24.66, \sigma = 6.82)$

Figure 7: Histogram of *Drosophila* NCNRs $(m = 5, j = 1, k = 0.25, \mu = 24.32, \sigma = 6.59)$ after randomly-shuffling

Figure 8: Histogram for CRMs, exons and NCNRs classified by $D(m = 5, j = 1)$

Figure 9: Histogram for CRMs, exons and NCNRs classified by $S_{50}(m = 5, j = 1)$

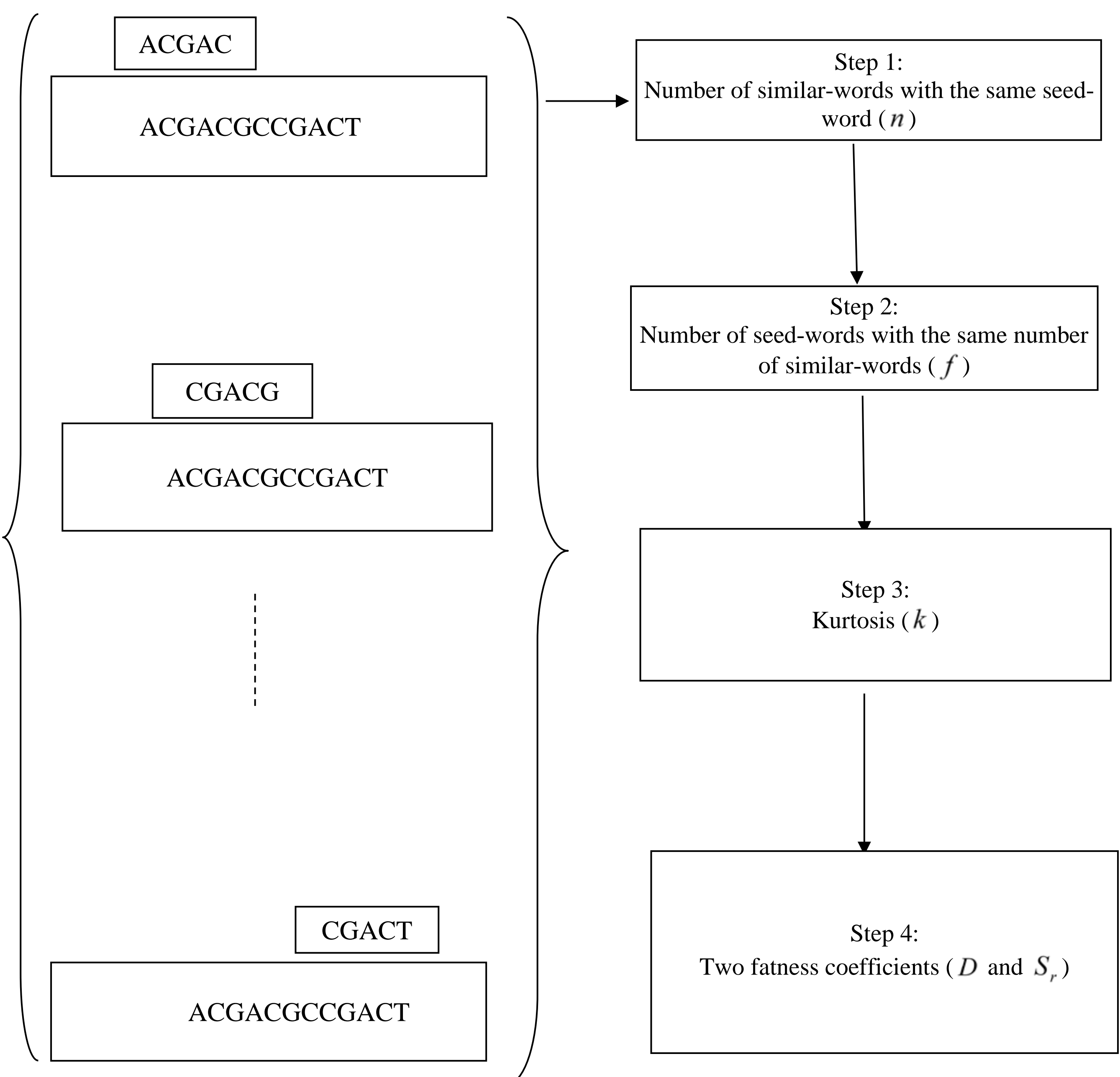

**Figure 1.** A flow chart of fat-tail test

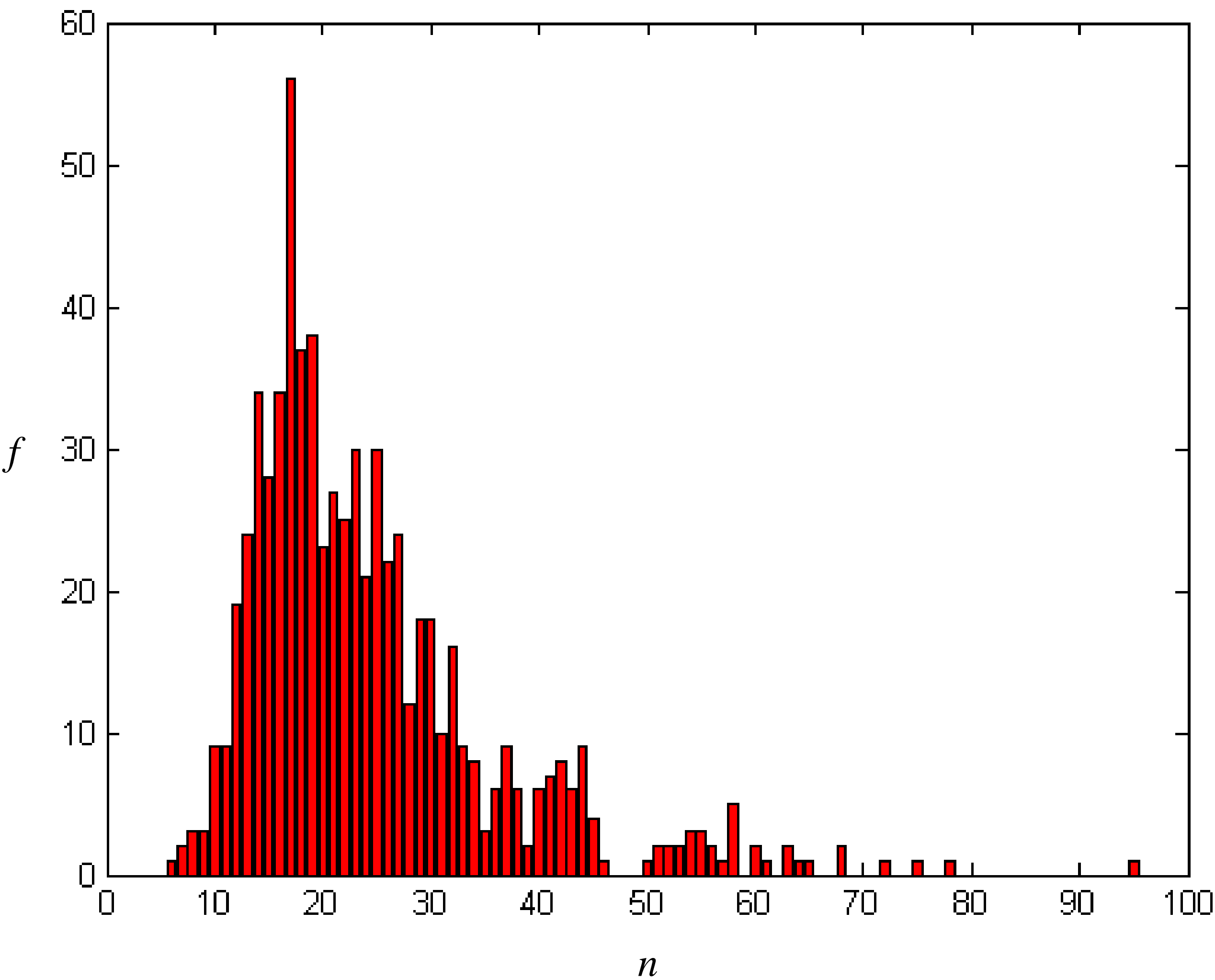

**Figure 2.** Histogram of *Drosophila* CRMs $\left(m = 5, j = 1, k = 4.19, \mu = 24.4, \sigma = 11.7\right)$

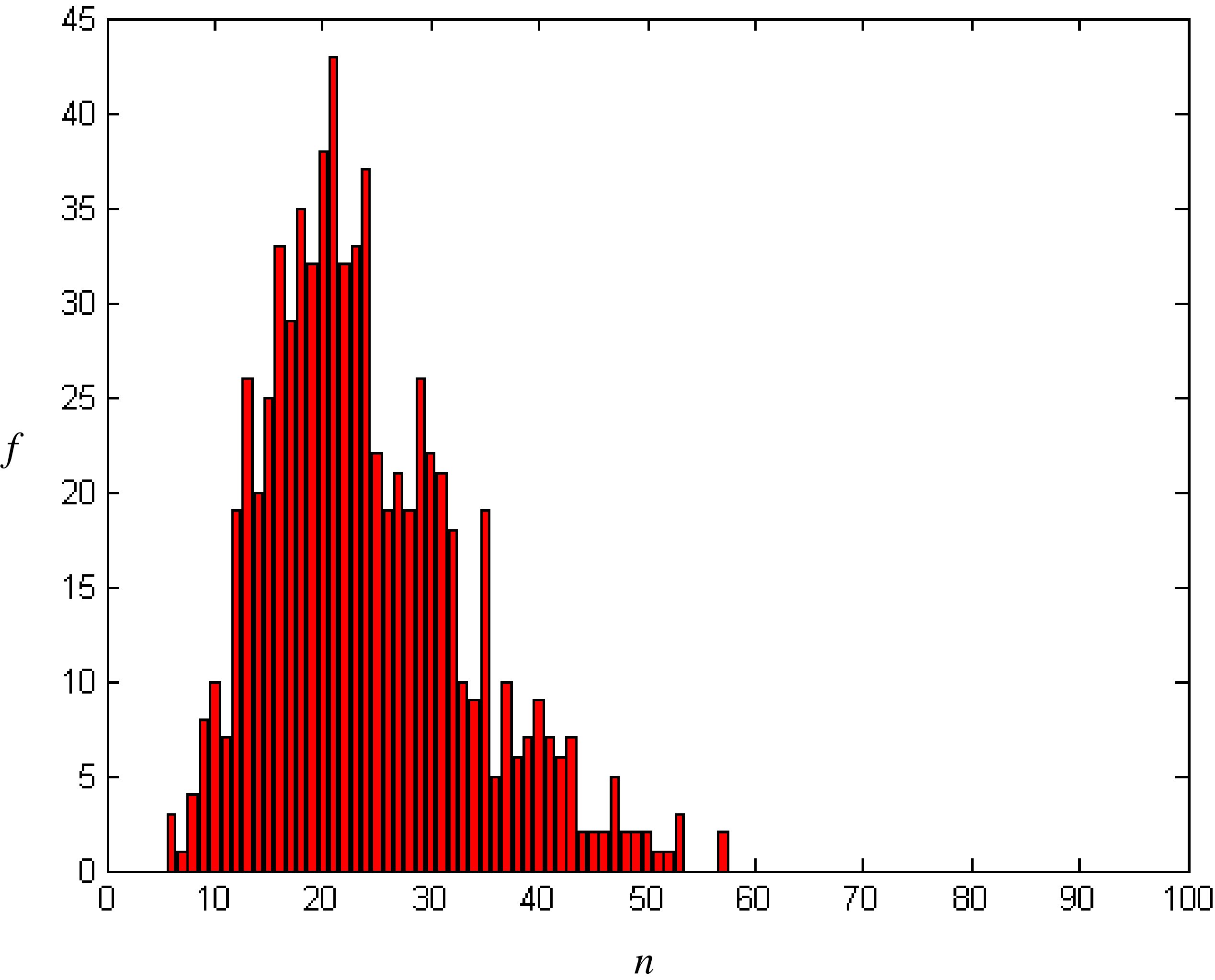

**Figure 3.** Histogram of *Drosophila* CRMs $\left(m = 5, j = 1, k = 0.19, \mu = 23.9, \sigma = 7.7\right)$ after randomly-shuffling

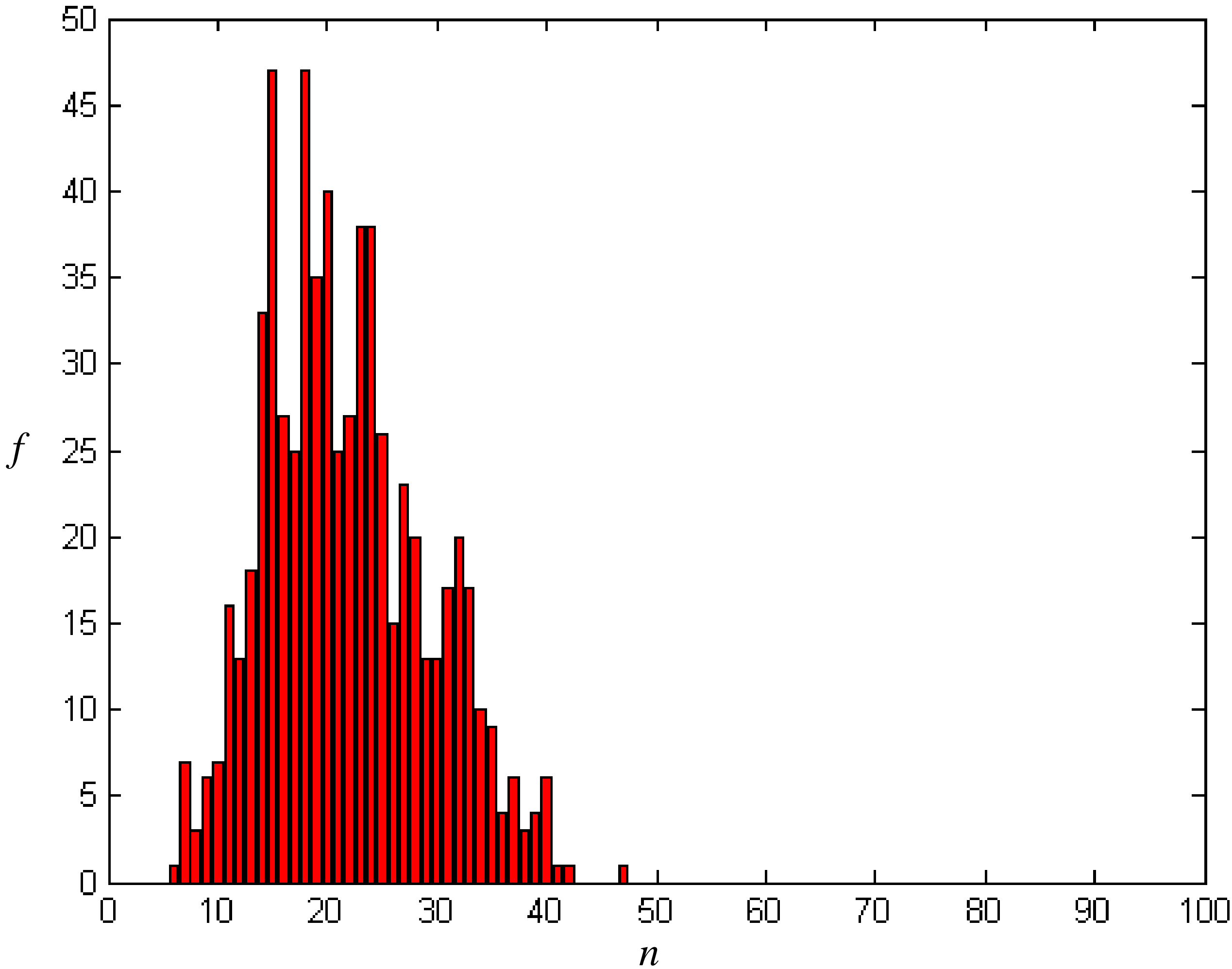

**Figure 4.** Histogram of *Drosophila* exons $\left(m = 5, j = 1, k = -0.28, \mu = 21.73, \sigma = 7.33\right)$

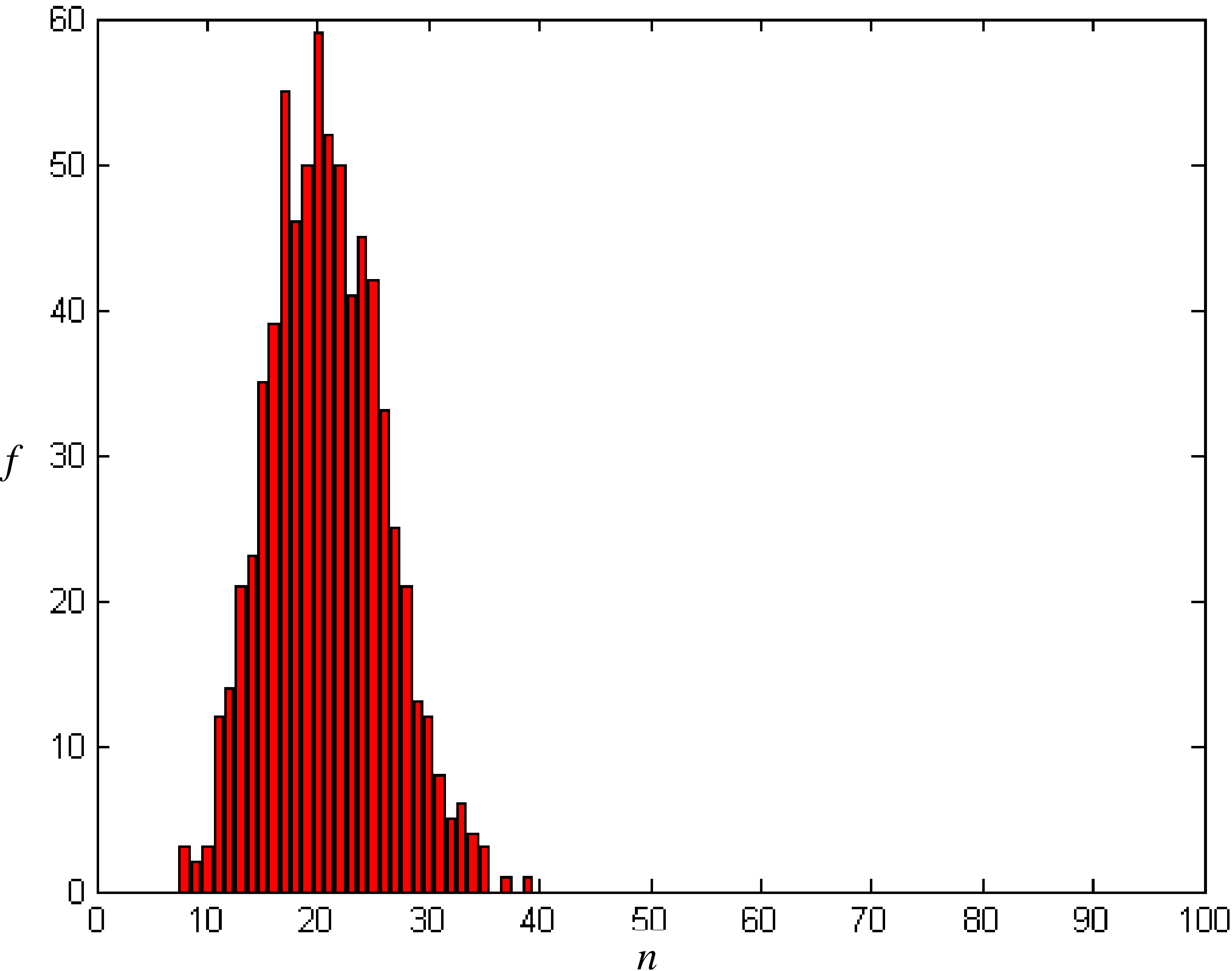

**Figure 5.** Histogram of *Drosophila* exons $\left(m = 5, j = 1, k = 0.35, \mu = 21.4, \sigma = 7.19\right)$ after randomly-shuffling

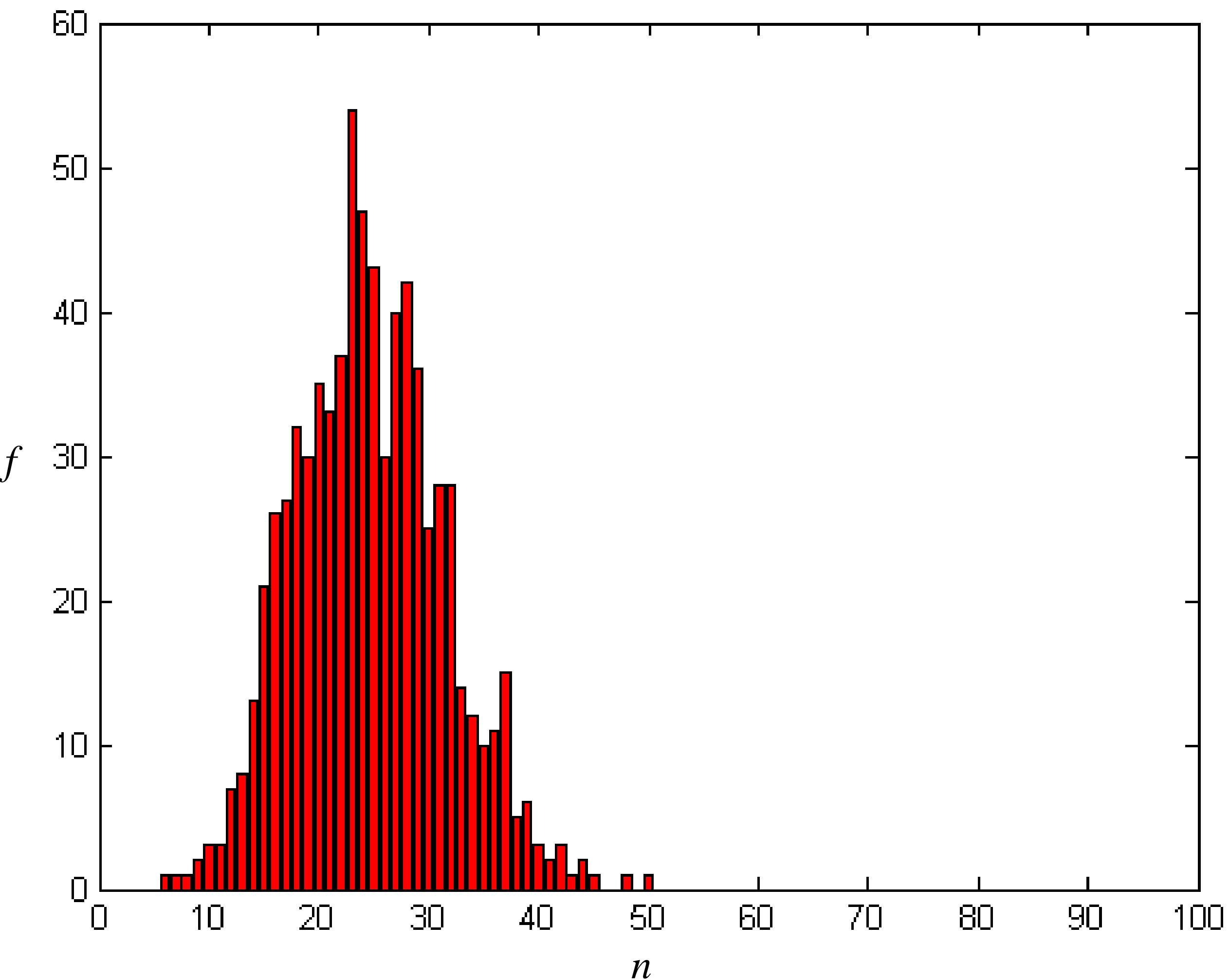

**Figure 6.** Histogram of *Drosophila* NCNRs $\left(m = 5,\ j = 1, k = 0.09, \mu = 24.66, \sigma = 6.82\right)$

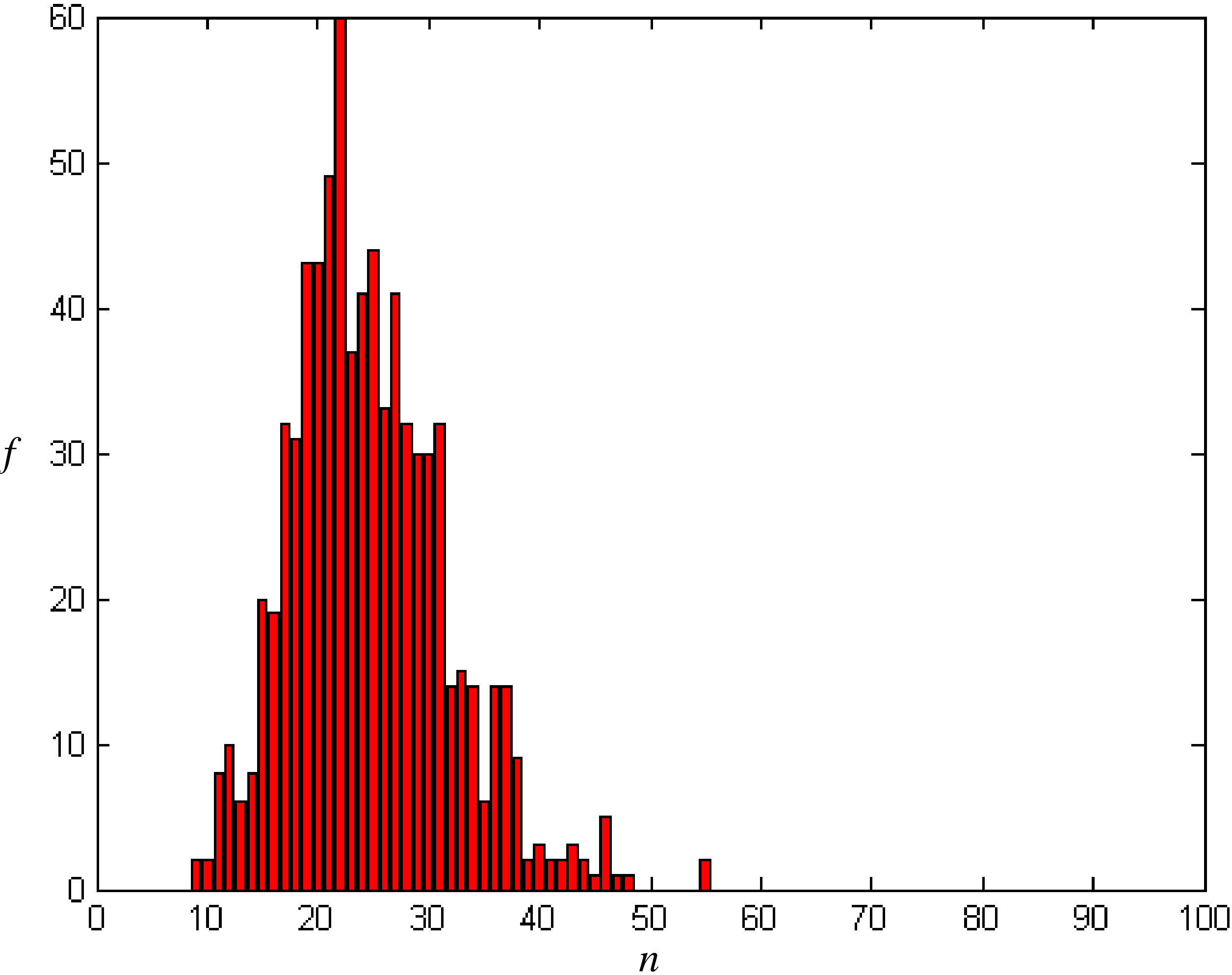

**Figure 7.** Histogram of *Drosophila* NCNRs $(m = 5, j = 1, k = 0.25, \mu = 24.32, \sigma = 6.59)$ after randomly-shuffling

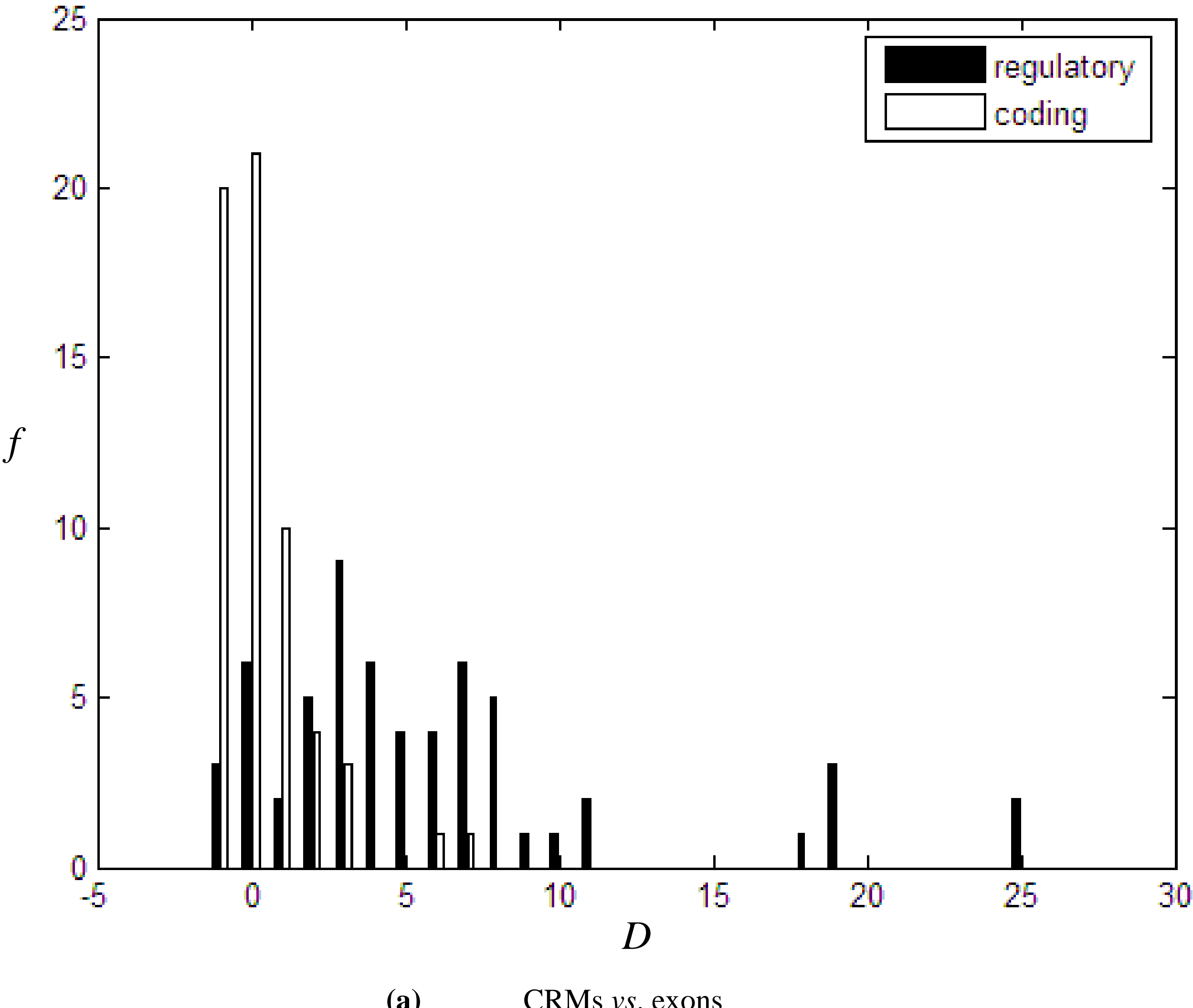

**(a)** CRMs *vs*. exons

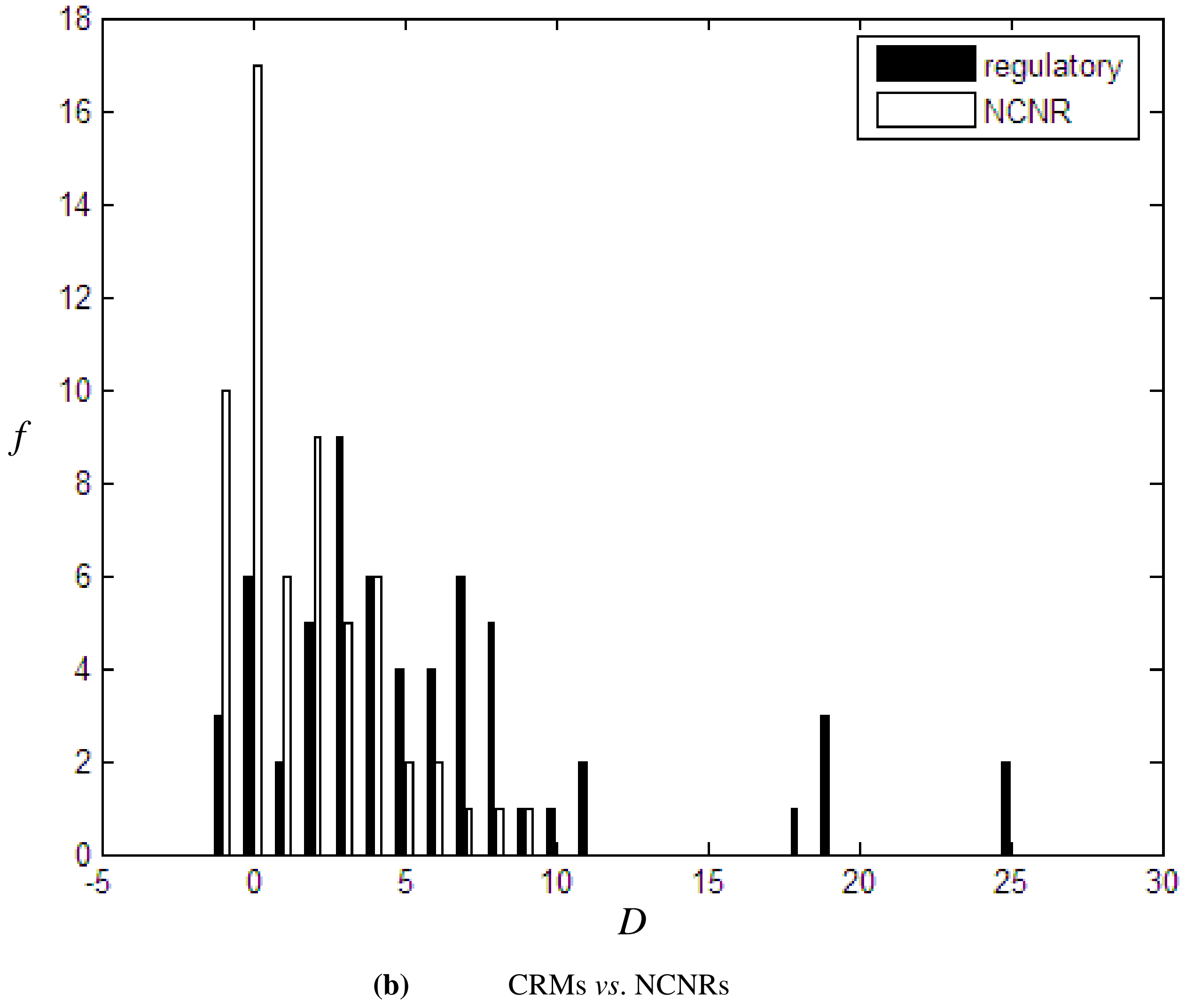

**(b)** CRMs *vs*. NCNRs

**Figure 8.** Histogram for CRMs, exons and NCNRs classified by $D(m = 5, j = 1)$

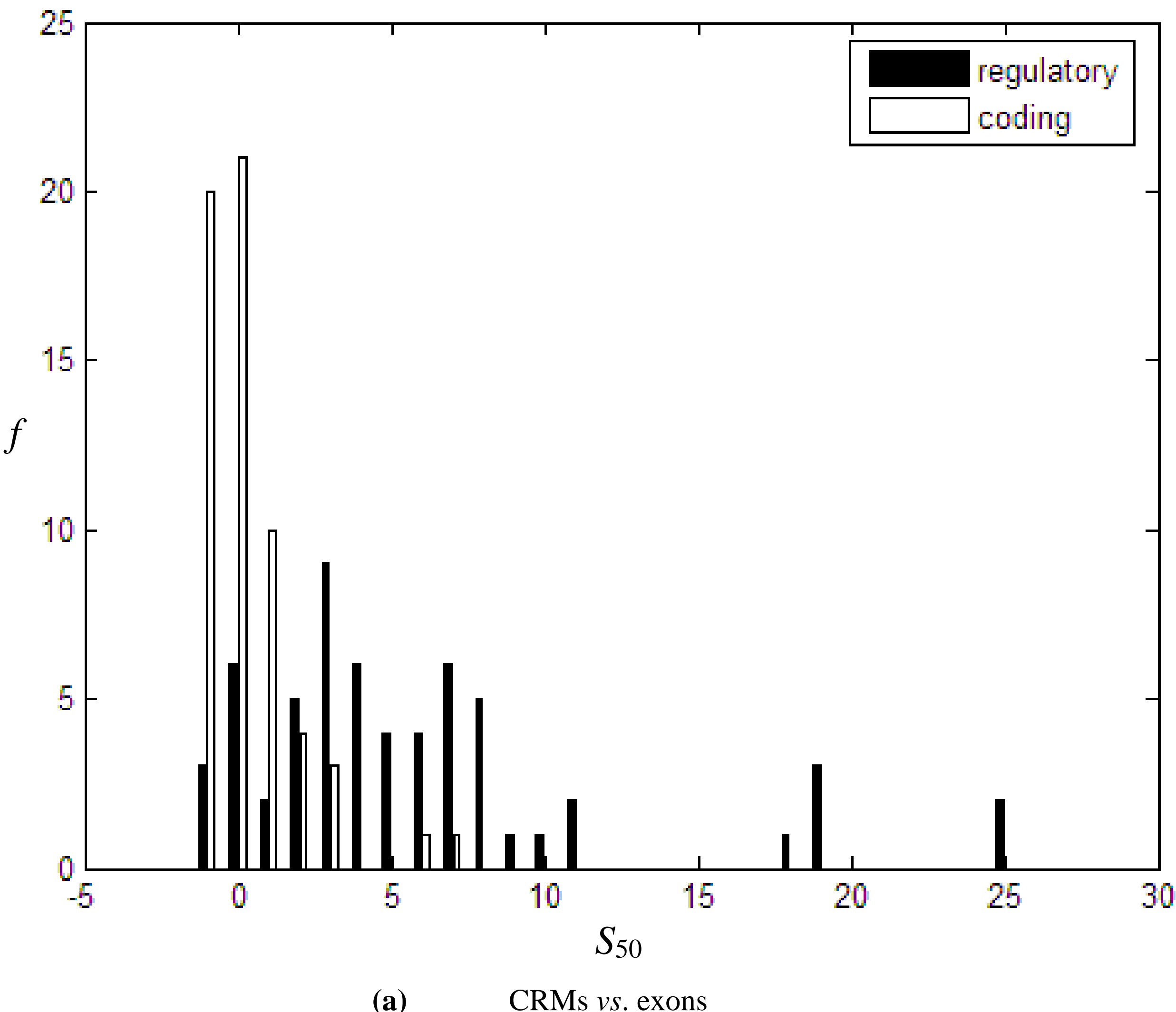

**(a)** CRMs *vs*. exons

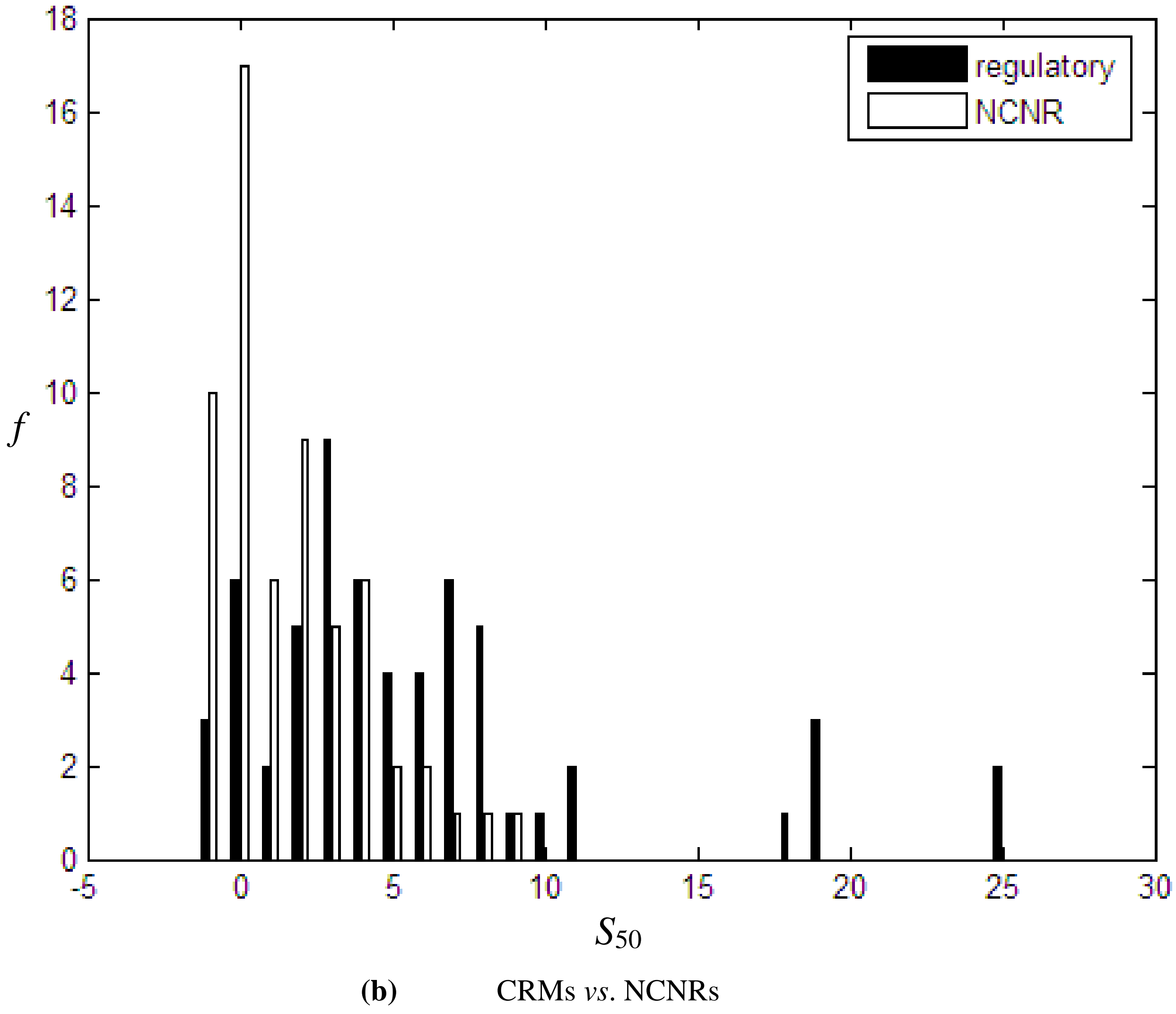

**(b)** CRMs *vs*. NCNRs

**Figure 9.** Histogram for CRMs, exons and NCNRs classified by $S_{50}\left(m = 5, j = 1\right)$